\documentclass[aps,prl,preprint,showpacs,preprintnumbers]{revtex4}

\usepackage{graphicx}
\usepackage{dcolumn}
\usepackage{bm}

\begin{document}

\title{Isotopic dependence of the giant monopole resonance in the
even-A $^{112-124}$Sn isotopes and the asymmetry term in nuclear
incompressibility}

\author{T.~Li$^{1}$, U.~Garg$^{1}$, Y.~Liu$^{1}$, R.~Marks$^{1}$,
B.K.~Nayak$^{1}$, P.V.~Madhusudhana~Rao$^{1}$, M.~Fujiwara$^{2}$,
H.~Hashimoto$^{2}$, K.~Kawase$^{2}$, K.~Nakanishi$^{2}$, S.~Okumura$^{2}$,
M.~Yosoi$^{2}$,  M.~Itoh$^{3}$, M.~Ichikawa$^{3}$, R.~Matsuo$^{3}$,
T.~Terazono$^{3}$,~M. Uchida$^{4}$, T.~Kawabata$^{5}$, H.~Akimune$^{6}$, 
Y.~Iwao$^{7}$, T.~Murakami$^{7}$, H.~Sakaguchi$^{7}$, S.~Terashima$^{7}$,
Y.~Yasuda$^{7}$,  J.~Zenihiro$^{7}$,  M.N.~Harakeh$^{8}$\\
\vspace{5ex}}

\affiliation{ $^{1}$ Department of Physics, University of Notre
Dame, Notre Dame, IN 46556, USA\\
$^{2}$ Research Center for Nuclear Physics, Osaka 567-0047, Japan\\
$^{3}$ Cyclotron and Radioisotope Center, Tohuku University, Sendai 980-8578,
Japan\\
$^{4}$ Department of Physics, Tokyo Institute of Technology, Tokyo 152-8850,
Japan\\
$^{5}$ Center for Nuclear Study, University of Tokyo, Tokyo 113-0033, Japan\\
$^{6}$ Department of Physics, Konan University, Kobe 658-8501, Japan\\
$^{7}$ Department of Physics, Kyoto University, Kyoto 606-8502, Japan\\
$^{8}$ Kernfysisch Versneller Instituut, University of Groningen,
9747 AA Groningen, The Netherlands}

\begin{abstract}

The strength distributions of the giant monopole resonance (GMR)
have been measured in the even-A Sn isotopes (A=112--124) with
inelastic scattering of 400-MeV $\alpha$ particles in the angular range
 $0^\circ$--$8.5^\circ$. We find that the experimentally-observed GMR
energies of the Sn isotopes are lower than the values
predicted by theoretical calculations that reproduce the GMR
energies in $^{208}$Pb and $^{90}$Zr very well. From the GMR data, a
value of $K_{\tau} = -550 \pm 100$ MeV is obtained for the
asymmetry-term in the nuclear incompressibility.

\end{abstract}

\date{\today}
\maketitle

Incompressibility of nuclear matter remains a focus of experimental
and theoretical investigations because of its fundamental importance
in defining the equation of state (EOS) for nuclear matter. The latter
describes a number of interesting phenomena from collective
excitations of nuclei to supernova explosions and radii of neutron
stars \cite{nkg}. The Giant Monopole Resonance (GMR) provides
a direct means to experimentally determine the nuclear incompressibility.

Experimental identification of the GMR
requires inelastic scattering of an isoscalar
particle--the $\alpha$ particle, for example--at extremely forward
angles, including 0$^{\circ}$, where the cross section for exciting
the GMR is maximal. Such measurements have improved considerably over the years
and it is now possible to obtain inelastic spectra virtually free of all
instrumental background directly \cite{exp-ND-3} and in coincidence with
proton- and neutron-decay \cite{exp-ND-1-1}. In recent work, the GMR
strength distributions have been extracted in a number of nuclei from a
multipole-decomposition analysis (MDA) of such ``background-free'' spectra
\cite{exp-RCNP-3,exp-RCNP-2,exp-RCNP-5,exp-RCNP-1,exp-ND-2,exp-ND-5,exp-ND-3}.

The excitation energy of the GMR is expressed in the scaling model
\cite{th-SS-4} as:
\begin{equation}
E_{GMR}=\hbar\sqrt{\frac{K_A}{m<r^2>}}
\end{equation}
\noindent where m is the nucleon mass, $<r^2>$ is the ground-state
mean-square radius, and $K_{A}$, the incompressibility of the nucleus.
In order to determine the incompressibility of
infinite nuclear matter, $K_\infty$, from the experimental GMR
energies, one builds a class of energy functionals, $E(\rho)$, with
different parameters which allow calculations for  nuclear matter
and finite nuclei in the same theoretical framework. The
parameter-set for a given class of energy functionals is
characterized by a specific value of $K_\infty$.
The GMR strength distributions are
obtained for different energy functionals in a self-consistent
RPA calculation. The $K_\infty$
associated with the interaction that best reproduces the GMR
energies is, then, considered the ``correct'' value. This procedure,
first proposed by Blaizot \cite{th-JPB-1}, is now accepted as the
best way to extract $K_\infty$ from the GMR data and,
following this procedure, it has been established that
both relativistic and non-relativistic
calculations are now in general agreement with $K_{\infty}$~=~240 $\pm$ 10 MeV
\cite{th-GC-3,th-JP-3,th-ss-2}.

The determination of the asymmetry term, $K_{\tau}$, associated with the
neutron excess (N-Z), remains very important because this term is crucial in
obtaining the radii of neutron stars in EOS calculations
\cite{exp-JML-1,latti,stein,exp-BAL-2}. Indeed, the radius of a neutron star
whose mass is between about 1 and 1.5 solar masses ($M_{\odot}$) is mostly
determined by the
density dependence of the symmetry-energy term \cite{exp-JML-2, th-JP-5}.
Previous attempts to extract this term
from experimental GMR data have resulted in widely different values, from
-320$\pm$180 MeV in Ref. \cite{sharma} to a range of -566$\pm$1350 MeV to
139$\pm$1617 MeV in Ref. \cite{shlomo2}.
Measurements of the nuclear incompressibility over a series of isotopes provide
a way to ``experimentally'' determine this asymmetry term in a direct manner.
The
Sn isotopes (A=112--124) afford such an opportunity since the asymmetry ratio,
((N-Z)/A), changes by more than 80\% over this mass range.

In this Letter, we report on new measurements on GMR in the even-A
Sn isotopes. The GMR has been identified previously in some
of the Sn isotopes as a compact peak in
measurements with inelastic $\alpha$-scattering \cite{exp-RCNP-1,
exp-TAM-3, exp-TAM-6, sharma} and 
although resonance parameters for GMR
in the Sn isotopes close to the values reported here have been extracted in
the past using less accurate techniques \cite{sharma}, the potentially large
systematic errors in those values necessitated the present measurements where
such problems have been eliminated. We find that the GMR energies in
the Sn isotopes are lower
than the values predicted in recent theoretical calculations even
though the interactions used in these calculations reproduce the GMR
energies in the ``standard'' nuclei, $^{208}$Pb and $^{90}$Zr, very
well. Also, we obtain a value $K_{\tau} = -550 \pm 100$ MeV from this data.

\begin{figure}[h]
\begin{center}
\includegraphics[width=8.6cm]{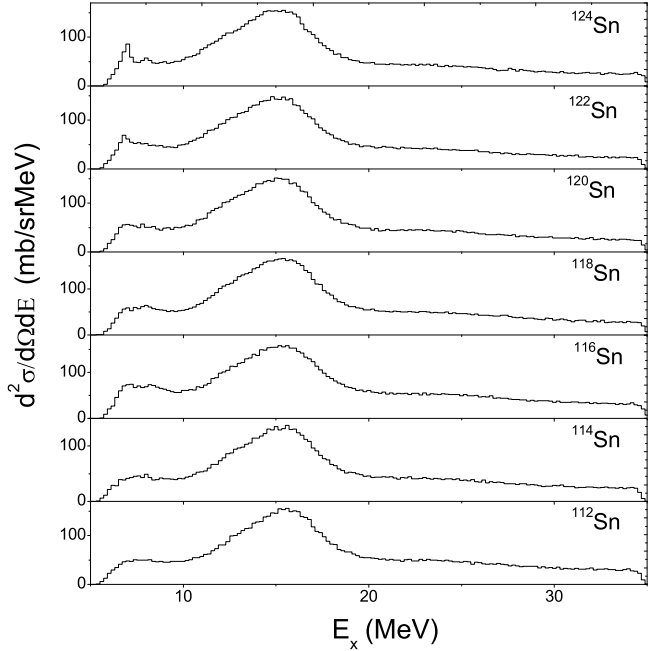}
\caption{\label{fig:sn0deg} Excitation-energy spectra for all even-A Sn 
isotopes, obtained from inelastic $\alpha$ scattering at
  $\theta_{lab}$ = 0.69$^\circ$.}
\end{center}
\end{figure}

The experiment was performed at the ring cyclotron facility of the
Research Center for Nuclear Physics (RCNP), Osaka University, using
inelastic scattering of 400-MeV $\alpha$ particles over
the angular range 0$^\circ$--8.5$^\circ$.
Details of the experimental
technique and the data analysis procedure have been provided
previously \cite{exp-ND-2, exp-RCNP-2, exp-RCNP-5} and are only
briefly described here. Inelastically-scattered $\alpha$ particles
were momentum-analyzed with the high-resolution magnetic
spectrometer ``Grand Raiden'' \cite{exp-RCNP-6} and detected in the
focal-plane detector system comprised of two multi-wire drift
chambers and two scintillators, providing particle identification as
well as the trajectories of the scattered particles.
The vertical position  spectrum obtained in
the double-focused mode of the spectrometer was exploited to
eliminate all instrumental background
\cite{exp-ND-2,exp-RCNP-2,exp-RCNP-5}. The background-free
``$0^\circ$'' inelastic spectra for the Sn isotopes are presented in
Fig.~\ref{fig:sn0deg}. In all cases, the spectrum is dominated by
the GMR peak near $E_x \sim 15$ MeV.

In order to extract the GMR strengths,
we have employed the now standard MDA procedure
\cite{exp-bb-1}. The cross-section
data were binned into 1-MeV energy intervals
between 8.5--31.5 MeV and for each
excitation energy bin,
the experimental 17-point angular distribution
$\frac{d\sigma^{exp}}{d\Omega}(\theta_{cm},E_x)$ was fitted by
means of the least-square method with the linear combination of
calculated distributions
$\frac{d\sigma^{cal}_{L}}{d\Omega}(\theta_{cm},E_x)$, so that: 
\begin{equation}
\frac{d\sigma^{exp}}{d\Omega}(\theta_{cm},E_x)=\sum_{L=0}^{7}\alpha_L(E_x)
\times \frac{d\sigma^{cal}_{L}}{d\Omega}(\theta_{cm},E_x)
\end{equation}
\noindent where  $\frac{d\sigma^{cal}_{L}}{d\Omega}(\theta_{cm},E_x)$ is
the calculated distorted-wave Born approximation (DWBA) cross
section corresponding to 100\% energy-weighted sum-sure (EWSR) for
the $L$-th multipole. This procedure provides strength distributions
simultaneously for various multipoles.

The DWBA calculations were performed following the method of
Satchler and Khoa \cite{th-GRS-1} using density-dependent single folding model,
with a Gaussian $\alpha$-nucleon potential for the real
part, and a Woods-Saxon imaginary term.
We used the transition densities and sum rules for
various multipolarities as described in Ref. \cite{book-2}.
The optical model (OM) parameters were obtained from analysis of
elastic scattering cross sections measured in a companion
experiment.

\begin{figure}
\includegraphics[width=8.6cm]{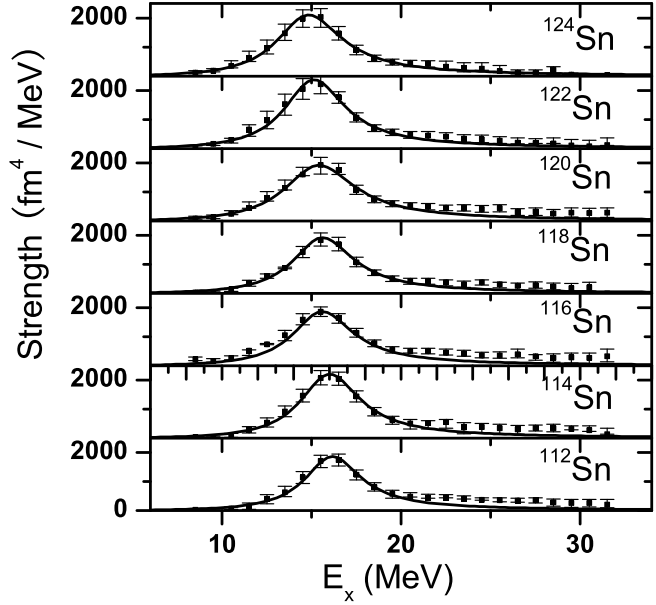}
\caption{\label{fig:str} GMR strength distributions obtained for the Sn isotopes
in the present experiment. Error bars
represent the uncertainty due to the fitting of the angular
distributions in MDA. The solid lines show Lorentzian fits to the
data.}
\end{figure}

Although all strength distributions up to $L$=3 have been reliably extracted
from the multipole decomposition, only the GMR strengths, the focus of this
paper, are shown in Fig.~\ref{fig:str}. The solid lines in the
figure represent Lorentzian fits to the observed strength
distributions. The choice of the Lorentzian
shape is arbitrary; the final results are not affected in any
significant way by using, instead, a Gaussian shape, for example. The finite
strength at the higher excitation energies is attributable to the mimicking
of $L$=0 angular distribution by components of the continuum
\cite{exp-RCNP-3, exp-ND-2}. 
The extracted GMR-peak parameters and the various moment ratios 
typically used in theoretical calculations
are presented in Table~\ref{tab:gmr}.

\begin{table*}
\caption{\label{tab:gmr}Lorentzian-fit parameters and various
  moment-ratios for the GMR strength distributions in the Sn
  isotopes, as extracted from MDA in the
  present work. $m_k$ is the $k$-th moment of the strength distribution:
$m_k = \int E^k_x S(E_x) dE_x$. All moment ratios have been calculated over
$E_x$~=~10.5--20.5 MeV. The errors quoted for EWSR are statistical only.
}
\begin{ruledtabular}

\begin{tabular}{ccccccc}

Target&$E_{GMR}$ (MeV)&$\Gamma$ (MeV)&EWSR &$m_1/m_0$

(MeV)&$\sqrt{m_3/m_1}$

 (MeV)&$\sqrt{m_1/m_{-1}}$ (MeV) \\

\hline $^{112}$Sn&$16.1\pm0.1$&$4.0\pm0.4$&$0.92\pm0.04$

&$16.2\pm0.1$&$16.7\pm0.2$&$16.1\pm0.1$ \\

$^{114}$Sn&$15.9\pm0.1$&$4.1\pm0.4$&$1.04\pm0.06$

&$16.1\pm0.1$&$16.5\pm0.2$&$15.9\pm0.1$ \\

$^{116}$Sn&$15.8\pm0.1$&$4.1\pm0.3$&$0.99\pm0.05$

&$15.8\pm0.1$&$16.3\pm0.2$&$15.7\pm0.1$ \\

$^{118}$Sn&$15.6\pm0.1$&$4.3\pm0.4$&$0.95\pm0.05$

&$15.8\pm0.1$&$16.3\pm0.1$&$15.6\pm0.1$ \\

$^{120}$Sn&$15.4\pm0.2$&$4.9\pm0.5$&$1.08\pm0.07$

&$15.7\pm0.1$&$16.2\pm0.2$&$15.5\pm0.1$ \\

$^{122}$Sn&$15.0\pm0.2$&$4.4\pm0.4$&$1.06\pm0.05$

&$15.4\pm0.1$&$15.9\pm0.2$&$15.2\pm0.1$ \\

$^{124}$Sn&$14.8\pm0.2$&$4.5\pm0.5$&$1.03\pm0.06$

&$15.3\pm0.1$&$15.8\pm0.1$&$15.1\pm0.1$ \\

\end{tabular}

\end{ruledtabular}

\end{table*}

\begin{figure}
\includegraphics[width=8.6cm]{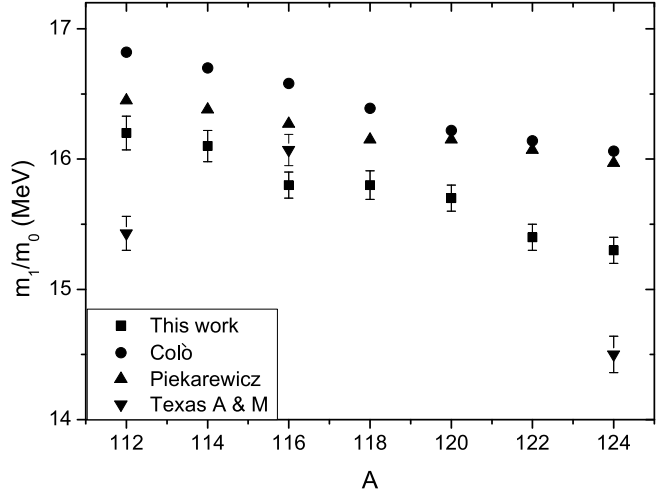}
\caption{\label{fig:expth} Systematics of the moment ratios
$m_1/m_0$
  for the GMR strength distributions in the Sn isotopes. The experimental
results (filled squares) are compared with results
  of non-relativistic RPA calculations by Col\`o \cite{th-GC-4} (filled circles)
and relativistic calculations of Piekarewicz \cite{th-JP-4} (triangles).
Results for $^{112}$Sn, $^{116}$Sn and $^{124}$Sn
  reported by the Texas A \& M group \cite{exp-TAM-6, exp-TAM-3} are also shown
  (inverse triangles). The differences between the present results and the
Texas A \& M results for $^{112,124}$Sn might be attributable to the background subtraction required in their analysis.}
\end{figure}

The moment ratios, $m_1/m_0$, for the GMR strengths in the Sn
isotopes are shown in Fig.~\ref{fig:expth}, and compared with recent
theoretical results from Col\`o (non-relativistic) \cite{th-GC-3,th-GC-4} and
Piekarewicz (relativistic)~\cite{th-JP-3,th-JP-4}. 
As can be seen, the calculations overestimate the
experimental GMR energies significantly (by almost 1 MeV in case
of the higher-A isotopes).
This is very surprising since the interactions used in these
calculations are those that very closely reproduce the GMR centroid energies
in $^{208}$Pb and $^{90}$Zr. Admittedly, there are uncertainties associated
with the range over which the experimental and theoretical distributions are
compared, and also with the assumptions inherent in the calculations regarding
widths. However, the calculations reported here are
identical in all respects to those performed for $^{208}$Pb and $^{90}$Zr, and
the experimental and theoretical centroids reported here have been calculated
over exactly the same excitation-energy range.
This disagreement remains a
challenge for the theory: Why are the tin isotopes so ``soft''?
Are there any nuclear structure effects that need to be taken into
account to describe the GMR energies in the Sn isotopes? 

The incompressibility of a nucleus, $K_{A}$, may be expressed as:
\begin{equation}
K_{A} \sim  K_{vol}(1 + cA^{-1/3}) + K_{\tau}((N - Z)/A)^{2} +
K_{Coul}Z^{2}A^{-4/3}
\end{equation}
\noindent
Here, $c \approx -$1 \cite{patra}, and $K_{Coul}$ is essentially
model-independent (in the sense that the deviations from one theoretical
model to another are quite small),
so that the associated term can be calculated for a given isotope.
Thus, for a series of
isotopes, the difference $K_{A}~-~K_{Coul}Z^{2}A^{-4/3}$ may
be approximated to have a quadratic relationship with the
asymmetry parameter, of the type $y = A + Bx^2$, with $K_{\tau}$
being the coefficient, $B$, of the quadratic term. It should be noted that
it has been established previously \cite{shlomo2,pears} that fits to the above
equation do not provide good constraints on the value of $K_{\infty}$.
However, this expression is being used here not to obtain a value for
$K_{\infty}$, but, rather, only to demonstrate the approximately
quadratic relationship between $K_{A}$ and the asymmetry parameter.

\begin{figure}[h]
\includegraphics[width=8.6cm]{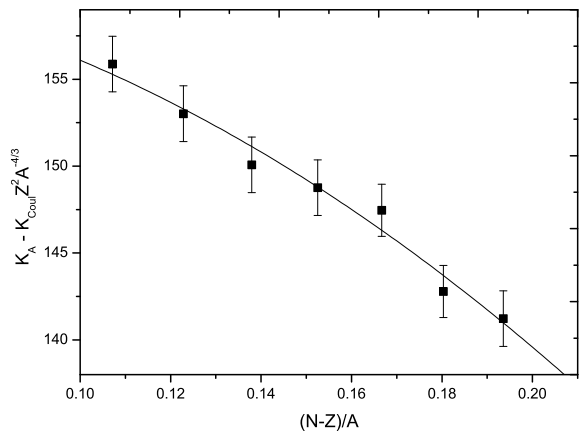}
\caption{\label{fig:sym} Systematics of the difference
$K_A - K_{Coul}Z^{2}A^{-4/3}$ in the Sn isotopes as a function of the
``asymmetry-parameter'' ((N-Z)/A);
$K_{Coul}$ = -5.2 MeV~\cite{sagawa}.
The solid line represents a
least-square quadratic fit to the data.}
\end{figure}

Fig.~\ref{fig:sym} shows the difference $K_A - K_{Coul}Z^{2}A^{-4/3}$
for the Sn isotopes investigated in this work
$vs.$ the asymmetry parameter, $((N-Z)/A)$.
The values of $K_A$ have been derived using the customary moment ratio
$\sqrt{m_{1}/m_{-1}}$ for energy of the GMR in Eq.~(1).
A quadratic fit to the data is
also shown. The fit gives $K_{\tau} = -$550 $\pm$ 40 MeV, with the
uncertainty attributed only to 
the fitting procedure. Including the uncertainties in $K_A$ in the fit adds 
another $\sim$25 MeV to this ``error'' (to $\pm$ 67 MeV) and the uncertainty in the value of $K_{Coul}$ ($\pm$ 0.7 MeV; see Ref.~\cite{sagawa}) would contribute $\sim$ 15 MeV.
Considering, further,
the approximation made in arriving at the quadratic expression,
the actual total uncertainty would be somewhat larger still; hence the rounded
value $K_{\tau} = -550 \pm 100$ MeV quoted earlier in the text.
This result is consistent with the value $K_{\tau} = -500 \pm 50$ MeV
obtained recently from an analysis of the isotopic transport ratios in
medium-energy heavy-ion reactions \cite{exp-BAL-1,exp-BAL-4}. As shown in
Ref. \cite{exp-BAL-2}, this value provides constraints on the radius
of a 1.4 $M_{\odot}$ neutron star that are in rather good
agreement with recent observational data. Thus, from the data on the
compressional-mode giant resonances, we now have ``experimental'' values of
both $K_{\infty}$ and $K_{\tau}$ which, together, can provide a means of
selecting the most appropriate of the interactions used in EOS calculations.
For example, this combination of values for $K_{\infty}$ and $K_{\tau}$
essentially rules out a vast majority of the Skyrme-type
interactions currently in use in nuclear structure calculations \cite{sagawa}.
A similar conclusion was reached for EOS equations in Ref.~\cite{exp-BAL-3}.
Furthermore, a more precise determination of $K_{\tau}$ provides
additional motivation for measurement of isoscalar monopole strength in
unstable nuclei, a focus of investigations at RIKEN and GANIL, for example
\cite{riken1, ganil1}.

In summary, we have measured the energies of the isoscalar giant
monopole resonance (GMR) in the even-A~$^{112-124}$Sn isotopes via
inelastic scattering of 400-MeV $\alpha$ particles at extremely
forward angles, including 0$^{\circ}$. The GMR energies are
significantly lower than those predicted for these isotopes by
recent calculations. Further, the asymmetry-term, $K_{\tau}$, in the
expression for the nuclear incompressibility has been determined to
be $-550 \pm 100$ MeV. 

We wish to express our gratitude to G. Col\`o and J. Piekarewicz
for providing results of their calculations prior to publication.
This work has been supported
in part by the National Science Foundation (Grants No. INT03-42942 and
PHY04-57120), and by
the Japan Society for the Promotion of Science (JSPS).

\bibliography{snbib}

\end{document}